\documentclass{osa-article}

\journal{osajournal}

\articletype{Research Article}

\begin{document}

\title{Chip-scale Full-Stokes Spectropolarimeter in Silicon Photonic Circuits}

\author{Zhongjin Lin, Tigran Dadalyan, Simon B\'elanger-de Villers, Tigran Galstian and Wei Shi\authormark{*}}

\address{Department of Electrical and Computer Engineering, Centre for Optics, Photonics and Laser (COPL), Universit\'e Laval, Qu\'ebec, QC G1V 0A6, Canada}

\email{\authormark{*}wei.shi@gel.ulaval.ca} 



\begin{abstract}
Wavelength-dependent polarization state of light carries crucial information about light-matter interactions. However, its measurement is limited to bulky, energy-consuming devices, which prohibits many modern, portable applications. Here, we propose and demonstrate a chip-scale spectropolarimeter implemented using a CMOS-compatible silicon photonics technology. Four compact Vernier microresonator spectrometers are monolithically integrated with a broadband polarimeter consisting of a 2D nanophotonic antenna and a polarimetric circuit to achieve full-Stokes spectropolarimetric analysis. The proposed device offers a solid-state spectropolarimetry solution with a small footprint of $1\times0.6~{\rm mm}^2$ and low power consumption of ${\rm 360~mW}$. Full-Stokes spectral detection across a broad spectral range of 50 nm with a resolution of 1~nm is demonstrated in characterizing a material possessing structural chirality. The proposed device may enable a broader application of spectropolarimetry in the fields ranging from biomedical diagnostics and chemical analysis to observational astronomy.
\end{abstract}

\section{Introduction}

Monitoring the spectrum (energy distribution over different wavelengths) and polarization state of light can provide key information about light-matter interactions, revealing nanostructures of a material \cite{mishchenko1995depolarization} and its chemical composition \cite{naumann1991microbiological}. Measurement of  wavelength-dependent state of polarization (or Stokes spectrum) is required for the studies of vibrational circular dichroism \cite{hemraz2014chiromers,kessler2018insight}, Rayleigh scattering \cite{wehner2001polarization}, vector magnetograph \cite{degl2006polarization}, and so on. A spectropolarimeter is used to measure the Stokes spectrum in these applications, which, in some fields, is also called spectroscopic ellipsometer \cite{fujiwara2007spectroscopic}. Among other applications, such devices play a critical role in the pharmaceutical industry for chiral separation and analysis of racemic drugs: many drugs are chiral compounds and marketed as racemates, whose chirality can be measured via circular dichroism \cite{nguyen2006chiral}. 

Over the last decade, the demand for compact, cost-effective, and low-power spectropolarimeters has increased dramatically. Recently, some miniature architectures of such devices have been demonstrated \cite{taniguchi2006miniaturized,chen2016integrated, okabe2009error, lee2017compressed}. However, these works were still exploring the possible improvements of traditional free-space optical components. Not surprisingly, this approach is limited to the decimeter scale footprint. A new paradigm is needed to reach qualitative changes. A chip-scale spectropolarimeter would be highly desired (particularly for portable, biomedical applications), which has yet to be reported. In this work, we will propose a chip-level spectropolarimeter in silicon photonic integrated circuits (PICs).

The proposed device includes four spectrometers and one polarimeter. Some high-performance on-chip spectrometers have recently been demonstrated on silicon PICs, such as Fourier transform spectrometer (FTS) \cite{zheng2019microring, souza2018fourier} and arrayed-waveguide grating spectrometer (AWGS) \cite{takahashi1990arrayed}. The silicon FTS typically consumes a significant power (at the watt scale \cite{souza2018fourier, zheng2019microring}) for the thermal tuning of waveguide delay. Such a high power raises concerns about reliability and scalability for a lab-on-a-chip system. The power consumption of AWGS is relatively low, but requires a large number of photodetectors, which complicates the measurement system and takes a large footprint \cite{cheben2007high}. The silicon FTS using arrayed Mach-Zehnder interferometers (MZIs) \cite{wang2019chip, velasco2013high} also has such a problem. The recently demonstrated digital silicon FTS using arrayed MZIs plus an on-chip thermo-optic switch fabric achieved a single photodetector solution with a lower power consumption, but at the cost of a larger footprint and increased control complexity \cite{kita2018high}.  

Here, we will propose a structure of serially coupled double microring resonator (SDMR) to realize the chip-level spectrometer. With the advantages of small size, high tunability, and low-power consumption, microrings (MRs) is an excellent choice for wavelength filter \cite{xia2011high}. But, the inverse relationship between the size and free spectral range (FSR) limits its application in spectroscopy. The structure of SDMR can well solve this problem. Because of the Vernier effect, the FSR of the SDMR can be largely extended without decreasing the diameter of the MRs. More importantly, the temperature required to cover the entire FSR can be drastically reduced.

Compared to spectrometers, integrated polarimeters have been much less investigated. Only a few full-Stokes polarimeters were demonstrated recently on a silicon chip \cite{lin2019chip,lin2018chip, wu2019fully,espinosa2017chip, martinez2018polarimetry}. The capacity of on-chip probing another dimension of photons than intensity and phase opened up immense opportunities for communications, quantum information, astronomy, and biomedical and chemical sensing. Our recent work \cite{lin2019optimal} has proven that an optimal polarimetric frame with a minimum number of photodetectors can be achieved in silicon PICs, offering excellent performance comparable to conventional free-space solutions but with significantly improved compactness and robustness. Nevertheless, despite their broadband operation, none of these integrated polarimeters can capture the wavelength dependence of Stokes parameters.  

In this paper, we propose and experimentally demonstrate, for the first time, a chip-scale spectropolarimeter in silicon PICs, which encompasses both functionalities of a full-Stokes polarimeter and a multichannel spectrometer. Analysis of an arbitrary state of polarization is realized using a 2D nanophotonic antenna and an on-chip interferometric circuit. With adopting SDMR, the proposed device simultaneously achieve a high resolution (1 nm) and a broad bandwidth (50 nm) in the Stokes spectrum. The efficacy of the proposed spectropolarimeter is demonstrated by characterizing the chirality of a cholesteric liquid crystal (CLC) slab. The whole device, including an array of photodetectors integrated on the same chip, takes a compact footprint of 0.6~mm$^2$ and a mean power consumption of only 360~mW.

\begin{figure} [ht]
\centering\includegraphics[width=110mm]{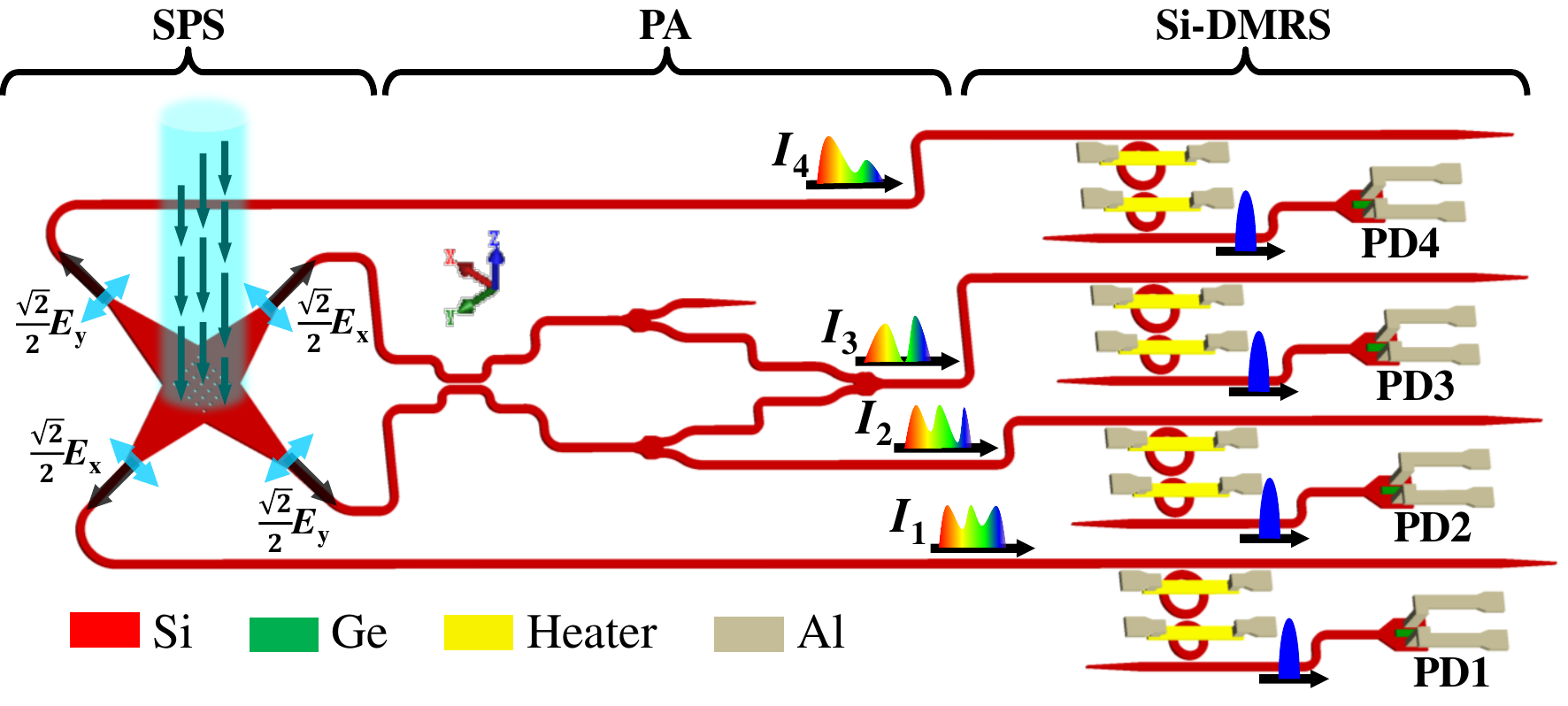}
\caption{Schematic of the proposed spectropolarimeter. The black arrows point to the propagating direction of light. SPS: surface polarization splitter; PA: polarization analyzer; Si-DMRS: our silicon dual-microring resonator sepectrometer; PDi: Ge photodetector of the i$^{th}$ Si-DMRS. }
\label{schematic_of_device}
\end{figure}

\section{Principle and design}
Figure~\ref{schematic_of_device} shows a schematic of the proposed device. It is designed based on a standard 220-nm-thick silicon-on-insulator (SOI) wafer with a 2~$\mu$m buried oxide layer and 3~$\mu$m oxide cladding. A surface polarization splitter (SPS) is used to project an arbitrary state of polarization into two orthogonal linearly polarized components (${E_x}$ and ${E_y}$) and couple them into difference waveguides.
A polarization analyzer (PA) in an interferometric circuit then converts the two orthogonal $E$-field components into four intensity channels. The spectrum of each intensity channel is measured using a spectrometer consisting of a thermally  tunable silicon dual-microring resonator and a Ge-PD. The four spectral measurements capture the full information of wavelength-dependent polarization, from which we can eventually retrieve the Stokes spectra of the input light via an matrix operation. 

The SPS makes use of a nanoantenna structure, consisting of a 2D  array of  sub-wavelength cylindrical holes on a Si substrate. The nanoantenna is designed so that both orthogonal linearly polarized components of the light, either from an optical fiber or free space, are coupled into the fundamental TE mode of the planar waveguides. Simultaneously, the SPS decomposes each orthogonal component equally into two paths in opposite directions as shown in Fig.~\ref{schematic_of_device}. More details about the design and performance of the SPS are given in \textbf{Appendix A}.

The PA circuit consists of a 3-dB broadband directional coupler (BDC) \cite{lu2015broadband}, three  Y-junctions \cite{lin2019broadband} for 3-dB power splitting/combination, and a few delay lines. Taking the outputs of the SPS, the PA  projects the Stokes vector of the incoming light into four intensity channels through interference operation: $I_1$ and $I_4$ from direct detection of $\frac{\sqrt{2}}{2}{E_x} $ and $\frac{\sqrt{2}}{2}{E_y}$, respectively; $I_2$ from the interference between $\frac{i}{2}{E_x}$ and $\frac{1}{2}{E_y}$; $I_3 $ from the interference between $\frac{1+i}{4}{E_x}$ and $\frac{1+i}{4}{E_y}$. Here, we denote the incoming polarization by a wavelength ($\lambda$) dependent Stokes vector: $\mathbf{S}\left(\lambda \right)=(S_0\left(\lambda \right),S_1\left(\lambda \right),S_2\left(\lambda \right),S_3\left(\lambda \right))^T$, where $\left(\star\right)^T$ means  the  transpose of  the matrix $\left(\star\right)$. Defining a wavelength-dependent intensity vector: $\mathbf{I}\left(\lambda \right)=(I_1\left(\lambda \right),I_2\left(\lambda \right),I_3\left(\lambda \right),I_4\left(\lambda \right))^T$, we find  the relationship between $\mathbf{I}\left(\lambda \right)$ and $\mathbf{S}\left(\lambda \right)$ given by,
\begin{equation}
\mathbf{S}\left(\lambda \right)\propto \mathbf{M_S}\left(\lambda\right)\cdot \mathbf{I}\left(\lambda \right)
\label{eq2-1}
\end{equation}
where $\mathbf{M_S}\left(\lambda\right)$ is the synthesis matrix of the PA. The BDC \cite{lu2015broadband} used in our design has a  wide bandwidth  in excess of 100 nm. Thus the synthesis matrix $\mathbf{M_S}\left(\lambda\right)$ is practically wavelength insensitive in the spectral range considered in this work and can be written by the following expression,
\begin{equation} 
 \mathbf{M_S}\left(\lambda\right)= 2\left( \begin{array}{cccc}
1 & 0 & 0 & 1 \\ 1 & 0 & 0 & -1 \\ -1 & 0 & 4 & -1 \\ -1 & 2 & 0 & -1
\end{array}  \right).
\label{eq2-2}
\end{equation}

\begin{figure} [ht]
\centering\includegraphics[width=100mm]{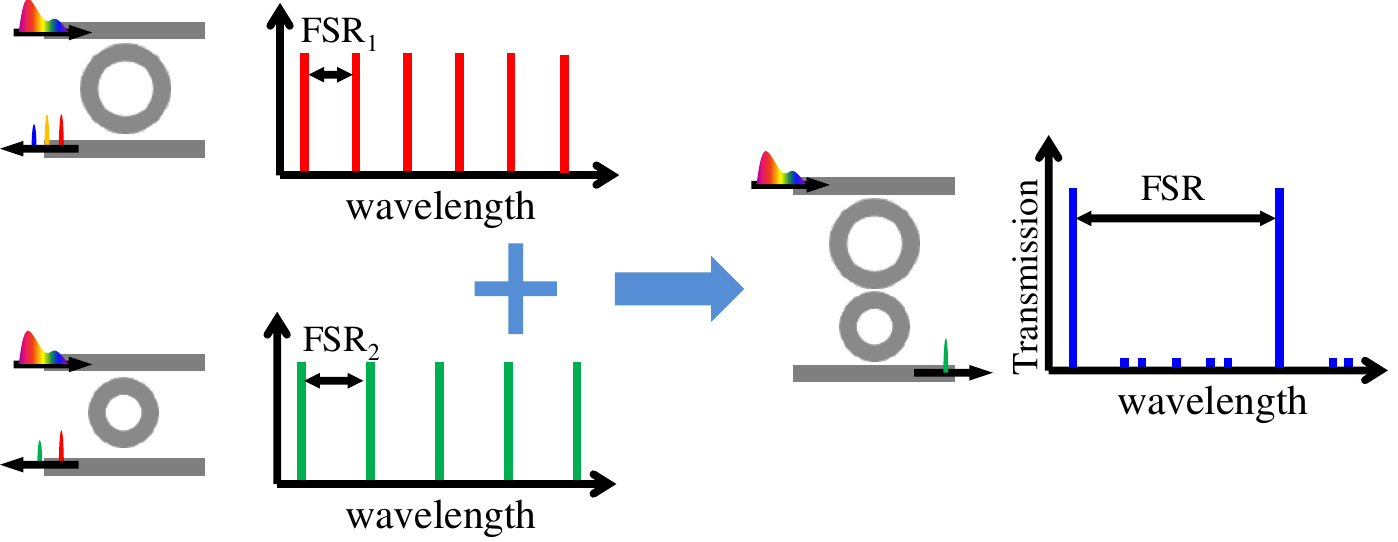}
\caption{Principle of the proposed Si-DMRS. FSR$_{1\left(2\right)}$ in the left side is the free spectral range of the single microring; the subscript 1(2) indicates the first (second) microring of the SDMR. FSR in the right side is the extended free spectral range of the SDMR.}
\label{Model_Dual_MR}
\end{figure}

Following the PA circuit, four silicon dual-microring resonator sepectrometers (Si-DMRSs) are used to measure the spectra of the intensity channels. Each Si-DMRS consists of an SDMR and a Ge-PD. The MRs in the SDMR have slightly different FSRs. Because of the Vernier effect, as shown in Fig. \ref{Model_Dual_MR}, the cascaded architecture can achieve a largely extended FSR without using ultra-small MRs which are challenging for fabrication on a wafer scale. The extended FSR of the SDMR is given by \cite{kolli2017design},
\begin{equation}
{\rm FSR}= \frac{{\rm FSR}_1\cdot {\rm FSR}_2}{|{\rm FSR}_1- {\rm FSR}_2|}=\frac{\lambda ^2}{\pi|D_1n_{g2}-D_2n_{g1}|}\approx \frac{\lambda ^2}{\pi n_{g1} |D_1-D_2|}
\label{eq2-3}
\end{equation}
where FSR$_{1\left(2\right)}$, $D_{1\left(2\right)}$, and $n_{g1\left(2\right)}$ are the FSR, diameter, and group index of the single MRs, respectively; the subscript 1(2) indicates the first (second) MR. We have $n_{g1}\approx n_{g2}$ when the diameters of the two rings are very close. According to Eq.~\ref{eq2-3}, we can increase the extended FSR of the SDMR by decreasing the difference of the diameters. A metal heater is used on the top of each MR to individually vary their temperatures. Tuning the heating powers (HPs) applied to the MRs, the wavelength of each intensity channel, $I_i\left(\lambda \right)$, can be continuously swept and then detected by a Ge-PD. More details about the analysis of the SDMR are shown in \textbf{Appendix B}. 

\section{Prototype}
\begin{figure} [ht]
\centering\includegraphics[width=110mm]{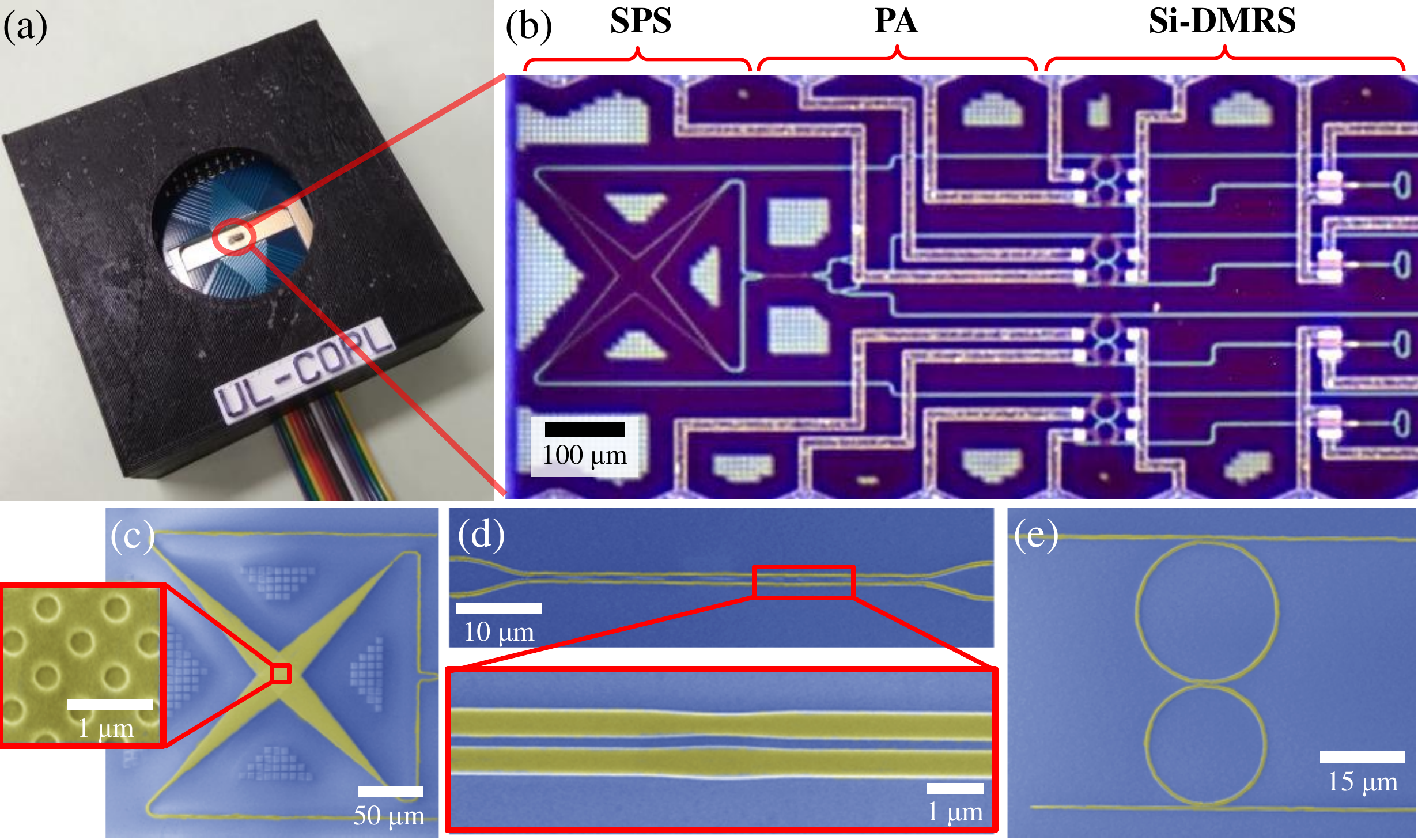}
\caption{Image of the fabricated spectropolarimeter. (a) the prototype of the fully packaged, plug-and-play spectropolarimeter with a ribbon cable for control and signal read-out. (b) The optical micrograph of the fabricated chip before being packaged. (c), (d) and (e) are the SEM images of the Si layer of the SPS, BDC, and SDMR, respectively. The inset in (d) presents the asymmetric-waveguide-based phase control section of the BDC for a broadband operation. SPS: surface polarization splitter; PA: polarization analyzer; Si-DMRS: our silicon dual-microring resonator sepectrometer.}
\label{photo_device}
\end{figure}

Figure~\ref{photo_device}(a) presents a packaged prototype of the proposed SP. Details of the fabrication and packaging processes are described in \textbf{Appendix F}. The fabricated silicon photonic chip sits in the center of the printed circuit board (PCB). Its footprint is $\sim 1\times0.6~$mm$^2$. The optical micrograph of the fabricated chip is depicted in Fig.~\ref{photo_device}(b). The chip includes 16 electric I/O ports. The connections between these I/O ports are presented in \textbf{Appendix D}. Figure \ref{photo_device}(c)-(e) present the scanning electron microscope (SEM) images of the SPS, BDC, and SDMR (silicon layer only), respectively. 

\section{Result}
\subsection{Si-DMRS performance} 
\begin{figure} [ht]
\centering\includegraphics[width=110mm]{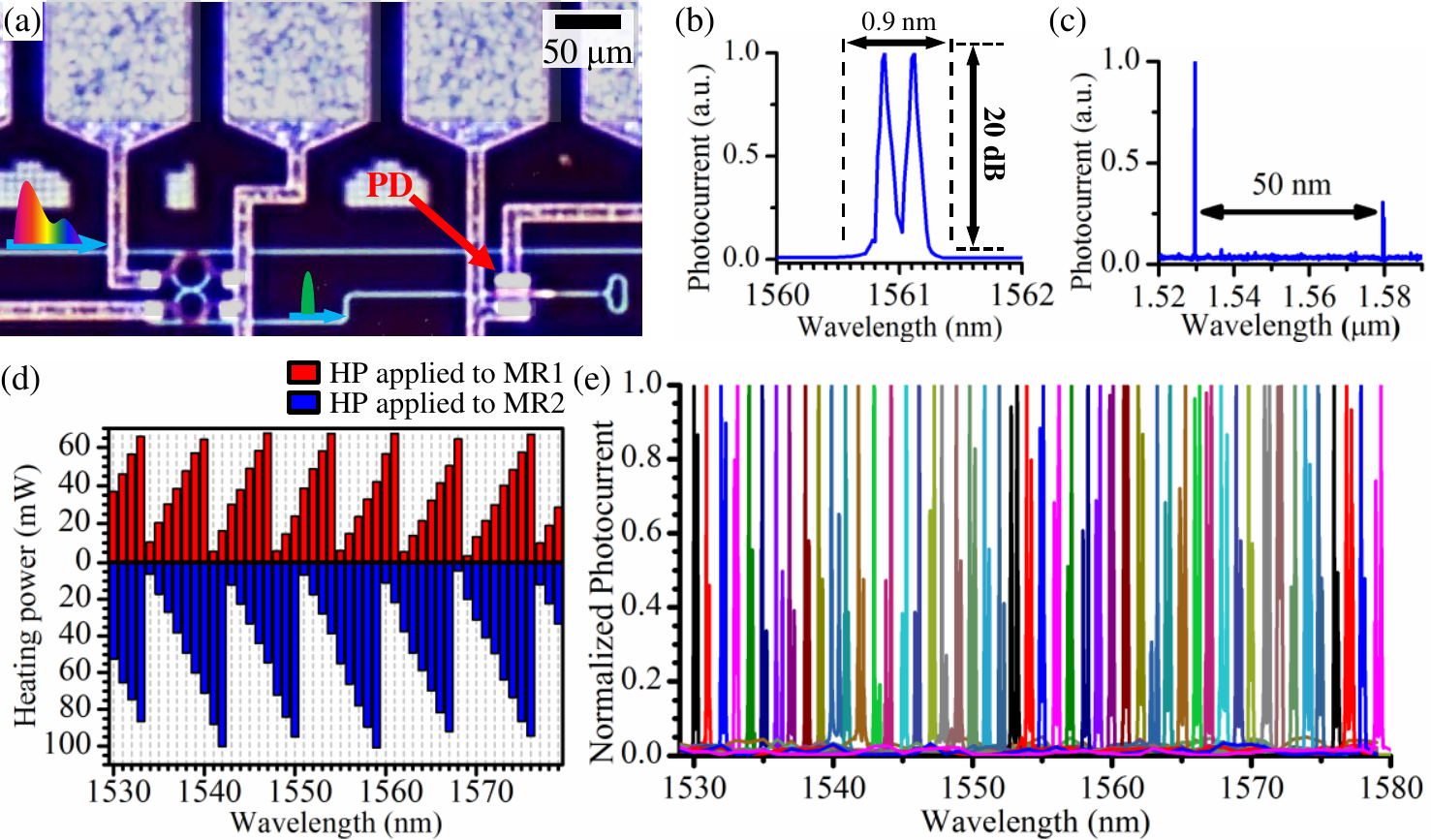}
\caption{Dual-MR characterization. (a) Optical micrography of a Si-DMRS. (b) and (c) The experimental transmission spectra from the drop port of the SDMR. (d) Relation between resonance wavelength and heating power on the heaters of MR1 (red dot) and MR2 (blue square). (e) The experimental transmission spectra of the drop port for the resonance wavelength from 1530~nm to 1579~nm. }
\label{Dual_MR}
\end{figure}

Before our experiment with the full-Stokes spectrometer, we firstly characterized a single Si-DMRS integrated with a Ge-PD on the same chip. Figure \ref{Dual_MR}(a) shows the optical micrography of the fabricated Si-DMRS. The diameters of two MRs are 26~$\mu$m (MR1) and 22~$\mu$m (MR2), respectively. A Ge-PD design without doped Ge or Ge-metal contacts \cite{zhang2014high} was adopted in our device. Because the germanium lattice is not disturbed by dopants or metal contacts, it allows for better performance in background loss, bandwidth, and dark current. The Ge-PD was measured to have a responsivity of 1.12~A/W and dark current of $\sim$15~nA at -4~V reverse bias, at 1550 nm wavelength. More information on the structure and performance of the Ge-PD is provided in \textbf{Appendix C}. Figure~\ref{Dual_MR}(b) shows the transmission spectrum from the drop port of the fabricated SDMR with a resonance wavelength near 1561 nm. A bimodal filter shape is observed and its 20-dB linewidth is near 0.9 nm. A sweeping step of 1~nm was chosen in the following experiment. An extended FSR of 50~nm is measured in Fig.~\ref{Dual_MR}(c).

The center wavelength as a function of HPs applied to MR1 and MR2 were calibrated for each channel using a tunable laser. The calibration result is shown in Fig.~\ref{Dual_MR}(d). The tuning efficiency is $\sim$10~mW/nm and $\sim$11~mW/nm for MR1 and MR2,  respectively. Thanks to the Vernier effect, the maximum HPs required to cover the entire extended FSR for MR1 and MR2 are only $\sim$70~mW and $\sim$100~mW, respectively. Figure~\ref{Dual_MR}(e) shows the measured transmission spectra of the Si-DMRS swept across the entire extended FSR (1530~nm to 1579~nm) with a step of 1 nm. 

\begin{figure} [ht]
\centering\includegraphics[width=70mm]{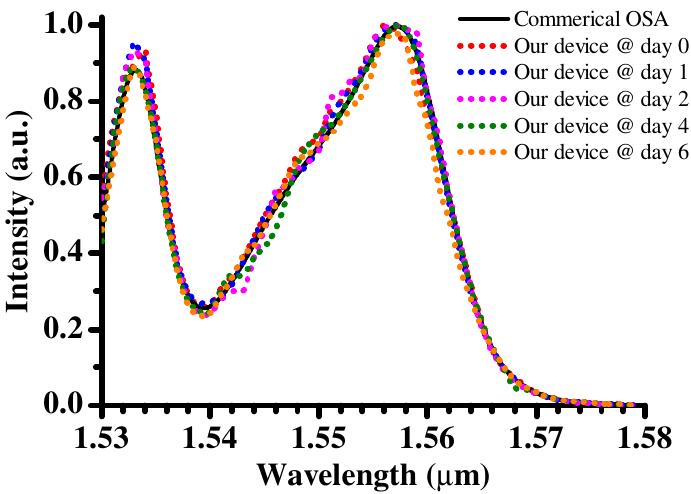}
\caption{Broadband spectrum reconstruction with the Si-DMRS. Solid black line is the spectrum recorded by a commerical OSA. The dotted lines are the measured results of the Si-DMRS over a week using the same calibration.}
\label{spectrum}
\end{figure}

Figure \ref{spectrum} compares the spectrum of a broadband source recorded by a commercial optical spectrum analyzer (OSA) as a reference (solid black) and that measured by the Si-DMRS (dotted lines), where a good agreement is observed. To verify the stability of the proposed device, we perform several measurements within a week using the same HP calibration; results are shown in Fig.~\ref{Dual_MR}(d). The measurement results show excellent agreement over six day, indicating a very stable operation of our device.

\subsection{Spectropolarimetric characterization of a chiral material}

\begin{figure} [ht]
\centering\includegraphics[width=110mm]{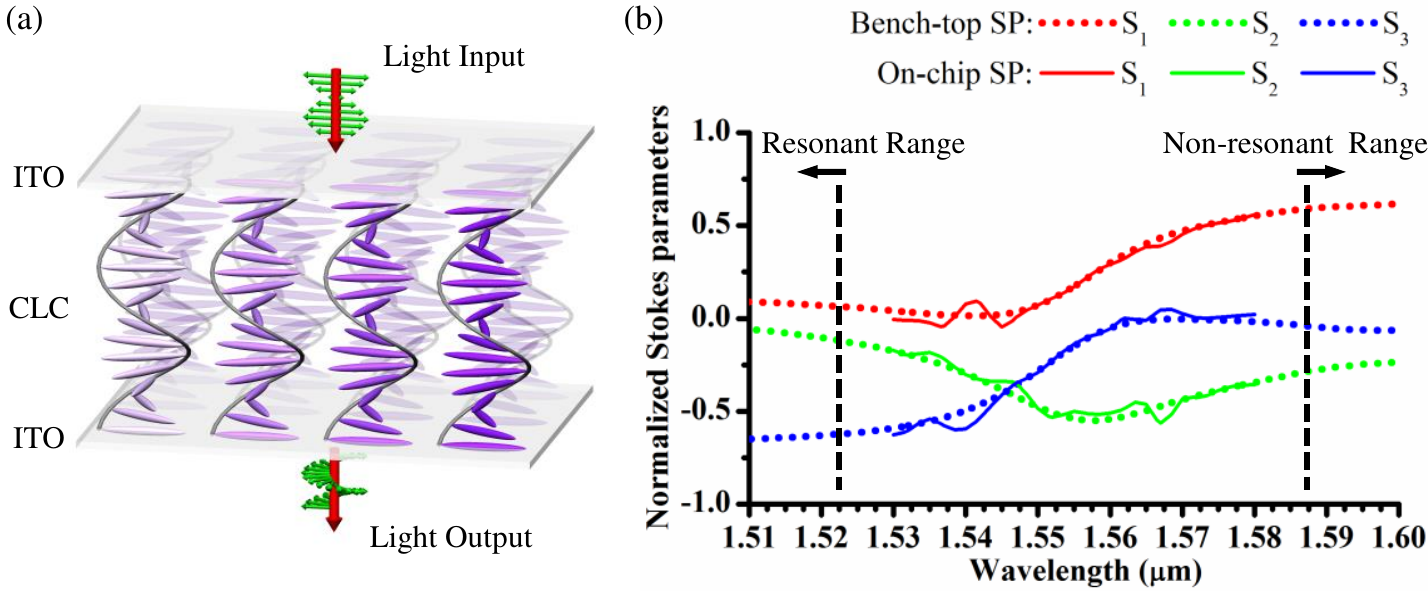}
\caption{On-chip spectropolarimeter characterization: (a) schematic of the CLC sample. (b)  Stokes spectra of the CLC sample, with a linear polarization input, measured by a commercial bench-top instrument (dotted lines) and our on-chip spectropolarimeter (solid lines). }
\label{SP_characterization}
\end{figure}

The spectropolarimeter's performance was tested  using a CLC slab \cite{de1995physics}. The schematic of the CLC slab is presented in Fig.~\ref{SP_characterization}(a). It consists of chiral molecules with a mechanical “twisting power”, which imposes a macroscopic helicoidal self-organization. As a result, the local average orientation of long molecular axis is periodically rotating from layer to layer,  forming a natural molecular helix (i.e. ``structural chirality''). With a proper choice of the molecular mixture parameters, the CLC slab  acts like a spectral “resonant” filter (e.g., Rocking filter) in a desired spectral range, which only left-handed (or right-handed) circular polarization can pass through. The most complex behavior occurs at the edges of the resonant wavelength range where polarization sensitive reflection and strong polarization rotations (along with strong dispersion) are present. To demonstrate efficacy of the proposed spectropolarimeter, we fabricated a CLC sample with an edge of the resonant range near 1550 nm. More details about the fabrication of CLC sample are given in \textbf{Appendix F}.

All the four Si-DMRSs were calibrated following the same procedure described in the previous section. The calibration results are depicted in \textbf{Appendix D}. The wavelength dependent synthesis matrix of the PA, $\bf M_s(\lambda)$, was also calibrated using four known independent polarization states. More details about the experiment are provided in \textbf{Appendix E}. Figure~\ref{SP_characterization}(b) shows the Stokes spectra after the CLC sample with a linear polarization input ($S_1 =1$). Excellent agreement is observed in the measurement results between our device (solid lines) and a commercial  bench-top instrument (dotted lines).  More details about the experiment in measuring CLC are provided in \textbf{Appendix E}. The resonant range of the fabricated CLC material is below $1.52~\mu$m. In the resonant range, only left-handed circular polarization can pass through the CLC; as shown in Figure~\ref{SP_characterization}(b), $S_3$ evolves towards -1, while $S_1$ and $S_2$ approaches zero as wavelength decreases.  While in the non-resonant range (beyond $1.58~\mu$m), the CLC material does not change the input polarization state. Therefore, as seen in Fig.~\ref{SP_characterization}(b), $S_1$ increases gradually from 0 towards 1 with the wavelength, while  $S_3$ increases from -1 to 0 in the non-resonant range.

\section{Discussion}
The entire spectropolarimeter, consisting of an SPS, a PA, and four spectrometers with Ge-PDs, has a compact footprint of $\sim 1\times0.6~$mm$^2$. In spite of compactness, our device remains a high performance with a high resolution (1 nm) and broad bandwidth (50 nm) of Stokes spectrum, which, however, has still not reached its limits. For example, according to Eq.~\ref{eq2-3}, we can obtain a bandwidth of 100~nm if increasing the diameter of MR$_2$ to $24~\mu$m. Besides, the spectral resolution of the proposed device can be further proved by another order of magnitude (to 0.1 nm) by decreasing the cross-coupling coefficient between the two MRs without introducing significant loss (as shown in \textbf{Appendix B}).

Because of the employment of the Si-DMRS, energy consumption is significantly reduced. Our spectropolarimeter only dissipates near 3.6~J of energy to complete one measurement of the Stokes spectra. Compared to traditional equipment, this value represents a few orders of magnitude improvement. Moreover, the energy consumption of the proposed spectropolarimeter can be significantly improved by adding thermal isolation trenches near the MRs ($>10$ times) \cite{ying2019thermally}, and by increasing the sweeping frequency of the HP ($>100$ times). Due to the limitation of our set-up, the sweeping frequency was only 5 Hz in our experiment. While the thermal response time of the MR is lower than 4~$\mu$s, indicating that a sweeping frequency of 250 kHz is possible \cite{atabaki2010optimization}. Assuming a higher sweeping frequency of 5 kHz for a larger number of spectral sweeping steps of 1,000 (versus 50 in our current experiment), the total energy consumption of the proposed spectropolarimeter is estimated to be only $\sim$72~mJ. In this case,  one measurement of Stokes spectra can be accomplished within 0.2~s.

\section{Conclusion}
Achieving an integrated spectropolarimeter on a silicon photonic chip paves the way towards fast, affordable full-Stokes spectroscopy. To decrease the cost and size of the device, traditional solutions come with a reduced number of spectroscopic components, and consequently, compromised measurement speed and Stokes spectral resolution. By contrast,  our solution allows for simultaneous achievement of a high speed and high resolution as all the Si-DMRSs can be integrated on a single chip with little increase in footprint and cost. Our device is fabricated using industry-standard silicon photonics foundry processes, indicating an easier path towards mass production using established large-wafer manufacturing facilities. The operating frequency range can be readily extended to the visible and mid-infrared regions by using other CMOS-compatible materials (e.g., SiN and Ge) but the same architecture. Leveraging the economies of scale and advantages of silicon PICs integration, the proposed spectropolarimeter has a vast potential for application in the fields of IoT, pharmaceutical analysis, astronomy, and so on.

\section{Appendix A: Surface Polarization Splitter}
\begin{figure} [ht]
\centering\includegraphics[width=120mm]{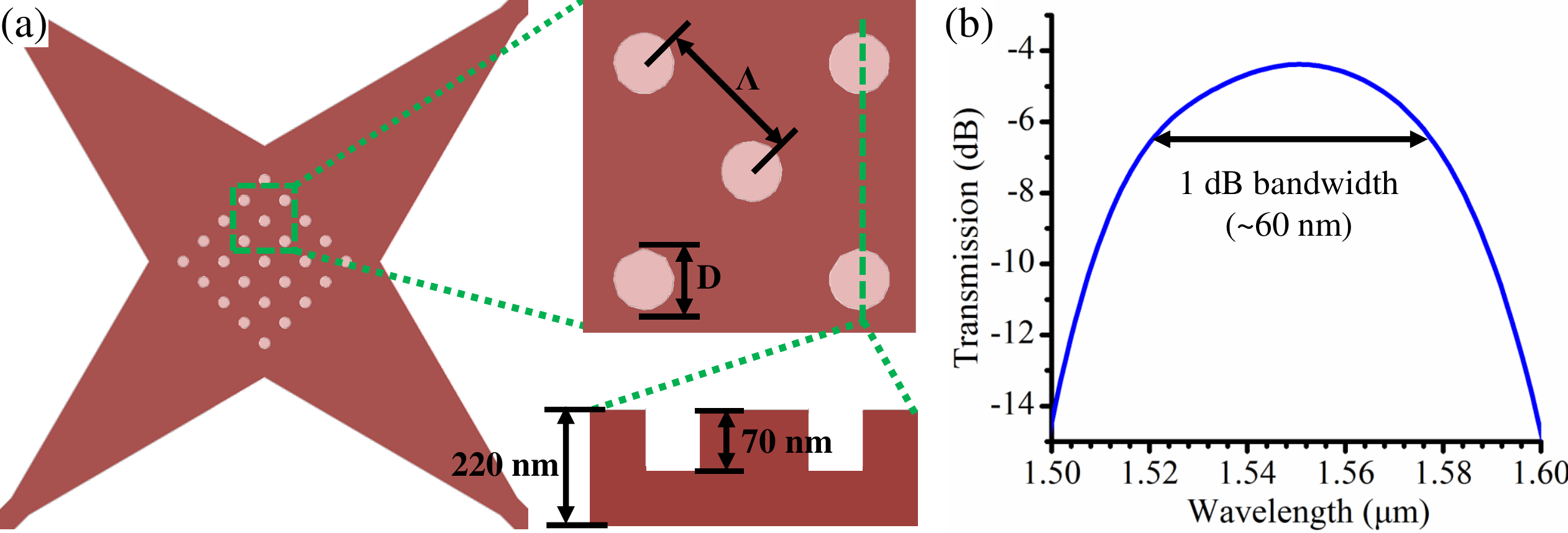}
\caption{(a) Schematic of SPS. The parameters $\Lambda$ and D are the period and diameter of the hole, respectively. (b) The simulated transmission spectra of SPS.}
\label{2DGC}
\end{figure}

The schematic of surface polarization splitter (SPS) is shown in Fig. \ref{2DGC}(a). The parameters $\Lambda$ and D are the period and diameter of the holes, respectively \cite{lin2019chip}. It is formed using a 40 $\times$ 40 array of cylindrical holes shallowly etched through silicon with an etched depth of 70 nm. The characteristics of SPS depend on $\Lambda$ and D. $\Lambda$= 540 nm and D = 280 nm have been chosen in our design. The design follows the same method given in our previous paper \cite{lin2019chip}. The simulation results of the proposed SPS are presented in Fig. \ref{2DGC}(b).

\section{Appendix B: Serially Coupled Double Microring Resonator}
\begin{figure} [ht]
\centering\includegraphics[width=120mm]{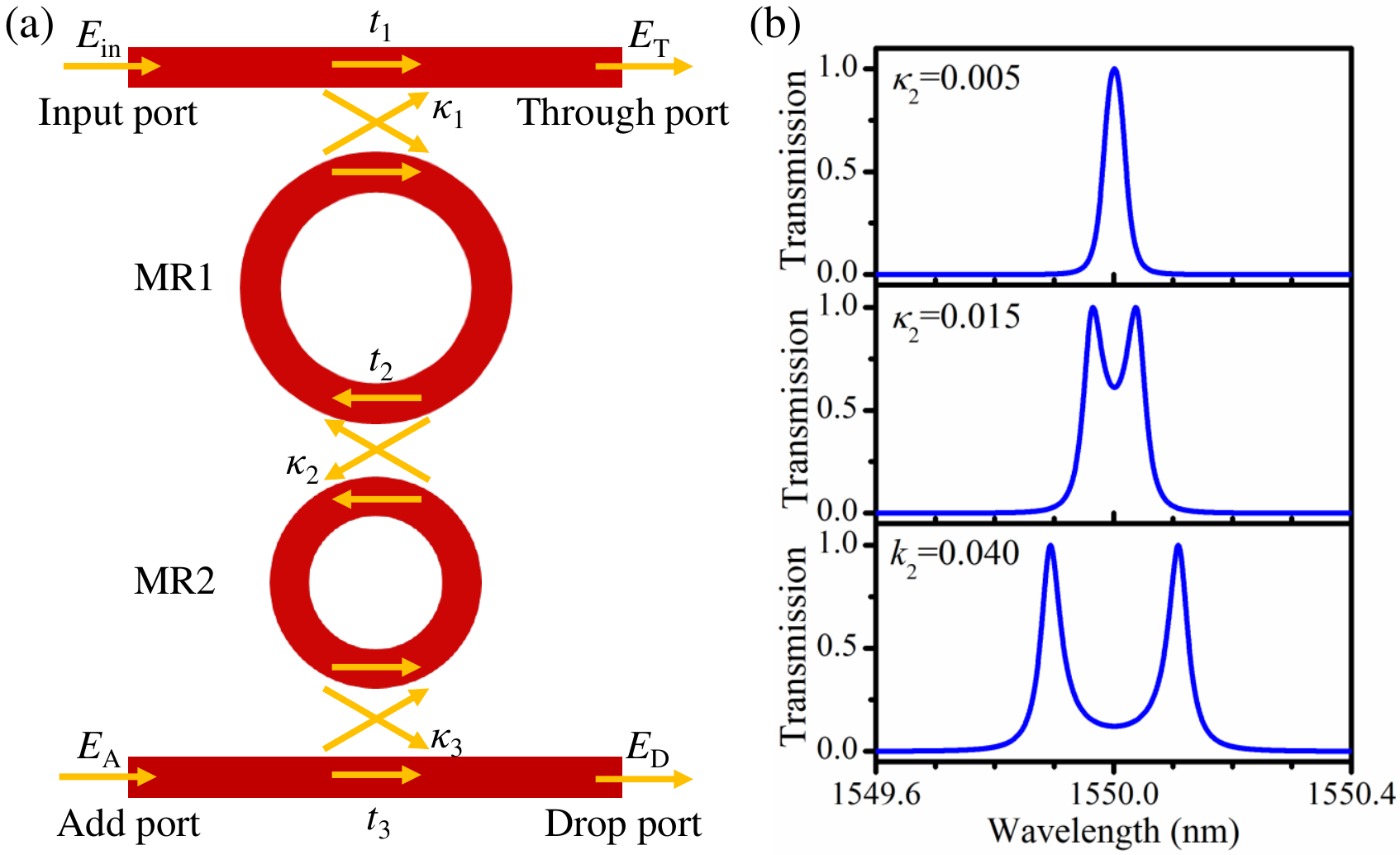}
\caption{(a) Schematic of SDMR. $t _{1,\left(2,~3\right)}$ and $\kappa _{1,\left(2,~3\right)}$ are normalized transmission and cross-coupling coefficients, respectively. (b) The simulated transmission spectra of the drop port in the case of $\kappa _2= 0.005$, $\kappa _2= 0.015$, and $\kappa _2= 0.04$ when the resonant wavelength is near 1550 nm.}
\label{SDMR_model}
\end{figure}

The schematic of the serially coupled double microring resonator (SDMR) is shown in Fig. \ref{SDMR_model}(a). $E_{in}$, $E_T$, $E_A$, and $E_D$ are the electrical fields in the input, through, add, and drop ports of the SDMR, respectively.  The relation between $\left(E_{in},~ E_T\right)^T$ and $\left(E_{A},~E_D\right)^T$ can be written by \cite{cho2008interferometric}, 
\begin{equation} 
 \left( \begin{array}{c}
E_{in} \\ E_{T} 
\end{array}  \right)=\left(\mathbf{C}_3 \mathbf{P}_2 \mathbf{C}_2 \mathbf{P}_1 \mathbf{C}_1 \right) \left( \begin{array}{c}
E_{A} \\ E_{D} 
\end{array}  \right)= \mathbf{M}\left(\begin{array}{c}
E_{A} \\ E_{D} 
\end{array}\right).
\label{eqB-1}
\end{equation}
where $\mathbf{C}_1$, $\mathbf{C}_2$, and $\mathbf{C}_3$ are the coupling matrices and can be given by,
\begin{equation} 
\mathbf{C}_{1,\left(2,~3\right)}=\frac{i}{\kappa _{1,\left(2,~3\right)}} \left( \begin{array}{cc}
t _{1,\left(2,~3\right)} & -1 \\ 1 & t _{1,\left(2,~3\right)}
\end{array}  \right)
\label{eqB-2}
\end{equation}
where $i^2=-1$. $t _{1,\left(2,~3\right)}$ and $\kappa _{1,\left(2,~3\right)}$ are the transmission and cross-coupling coefficients, respectively, assuming $|t _{1,\left(2,~3\right)}|^2+|\kappa _{1,\left(2,~3\right)}|^2=1$. $\mathbf{P}_1$, $\mathbf{P}_2$ are the propagation matrices, and can be given by,
\begin{equation} 
\mathbf{P}_{1,\left(2\right)}= \left( \begin{array}{cc}
0 & \sqrt{\alpha _{1,\left(2\right)}}\exp{\left(-i\pi \beta _{1,\left(2\right)}R_{1,\left(2\right)}\right)} \\ \sqrt{\alpha _{1,\left(2\right)}}\exp{\left(i\pi \beta _{1,\left(2\right)}R_{1,\left(2\right)}\right)} & 0
\end{array}  \right)
\label{eqB-3}
\end{equation}
where $\alpha _{1,\left(2\right)}$, $\beta _{1,\left(2\right)}$, and $R _{1,\left(2\right)}$ are the round trip attenuation, propagation constant, and radius of MR1(2), respectively. If we set $\mathbf{M}$ as,
\begin{equation} 
\mathbf{M}= \left( \begin{array}{cc}
m_{11} & m_{12} \\ m_{21} & m_{22}
\end{array}  \right)
\label{eqB-4}
\end{equation}
, the normalized output intensities at the “drop” ($E_D$) port and “through” ($E_T$) port of the SDMR can be obtained by,
\begin{equation} 
|E_D|^2=|-\frac{\text{Det}\left(\mathbf{M}\right)}{m_{12}}|^2,~|E_T|^2=|-\frac{m_{11}}{m_{12}}|^2,
\label{eqB-5}
\end{equation}
When $R_1$= 13 $\mu$m and $R_2$= 11 $\mu$m, $\kappa _1\approx 0.125$ and $\kappa _3\approx 0.115$. In this case, the spectra of the drop port as a function with the cross-coupling coefficient $\kappa _2$ are presented in Fig. \ref{SDMR_model}(b). We can observe that the 20-dB linewidth of the SDMR decreases with reducing the cross-coupling coefficient $\kappa _2$.  When $\kappa _2= 0.005$, the 20-dB linewidth of the SDMR is equal to 0.2 nm, indicating a resolution of 0.1 nm is available. According to the vernier effect, the free spectral range (FSR) of the SDMR can be given by \cite{kolli2017design},
\begin{equation}
{\rm FSR}= \frac{{\rm FSR}_1\cdot {\rm FSR}_2}{|{\rm FSR}_1- {\rm FSR}_2|}=\frac{\lambda ^2}{\pi|D_1n_{g2}-D_2n_{g1}|}\approx \frac{\lambda ^2}{\pi n_{g1} |D_1-D_2|}
\label{eqB-6}
\end{equation}
where FSR$_1$ and FSR$_2$ are the free spectral range of the MR1 and MR2, respectively. Equation \ref{eqB-6} shows that the FSR of the SDMR can be increased by decreasing the difference of the radius of two rings.

\section{Appendix C: Ge-on-Si photodetector}
\begin{figure} [ht]
\centering\includegraphics[width=130mm]{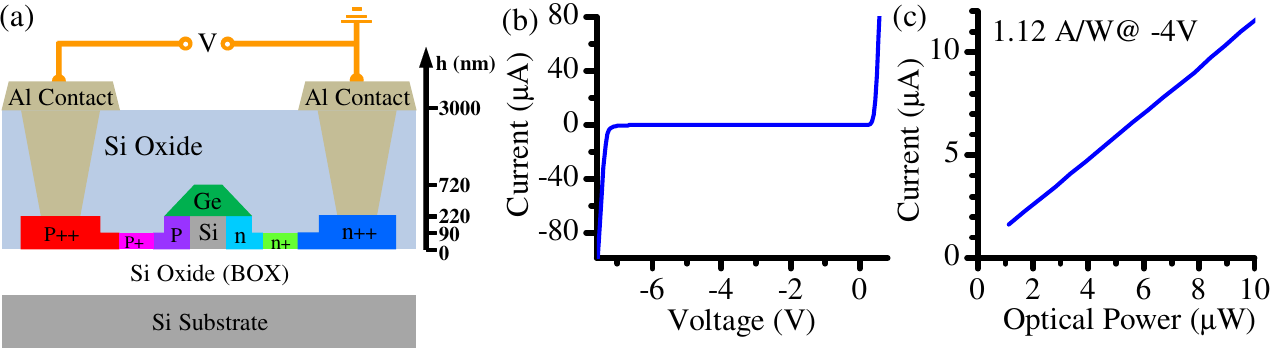}
\caption{(a) The cross-sectional schematic of the Ge-PD. (b) I-V curve in darkness. (c) Photocurrent as a function of optical power for the bias voltage of -4 V.}
\label{PD}
\end{figure}

The cross-sectional schematic of the Ge-PD is shown in Fig. \ref{PD}(a). The I-V curve in darkness is presented in Fig. \ref{PD}(b). The breakdown voltage situates at -7 V. Figure \ref{PD}(c) depicts the photocurrent of the Ge-PD as a function with the optical power when applying a bias voltage of -4 V.  In this case, the responsivity and dark current of the Ge-PD are near 1.12 A/W and 15 nA, respectively.

\begin{figure} [h!]
\centering\includegraphics[width=130mm]{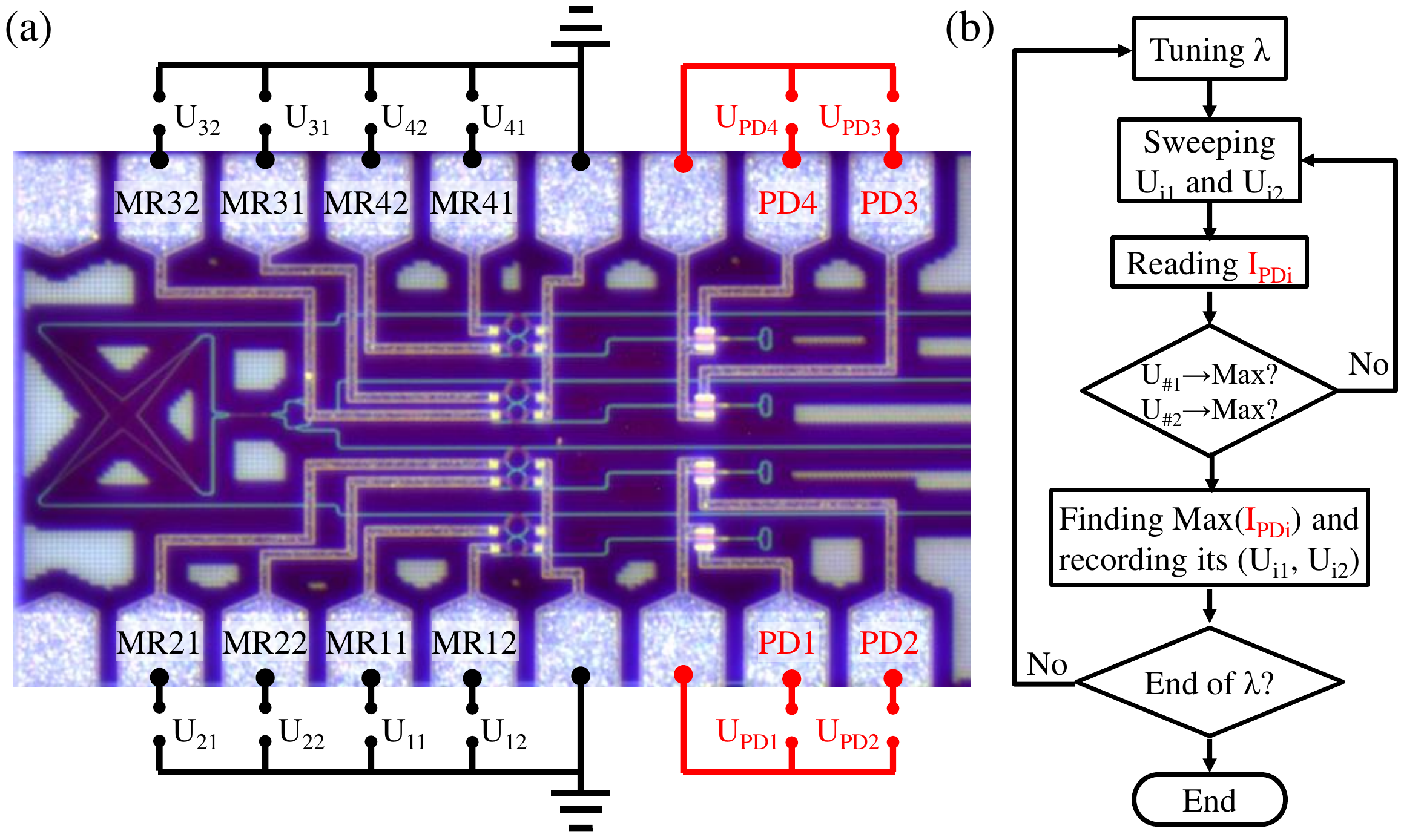}
\caption{ (a) The schematic of the electric connections. (b) The flowchart of searching the corresponding ($\text{U}_{i1}$, $\text{U}_{i2}$) for each wavelength. $\text{U}_{i1}$ and $\text{U}_{i2}$ are the voltages applied to 1st and 2nd MRs of the $i^{th}$ SDMR, respectively. PD$_i$ means the photodtector of the $i^{th}$ SDMR. I$_{\text{PD}i}$ is the current read from the PD$_i$.}
\label{HP_calibration_model}
\end{figure}

\section{Appendix D: Heating Power Calibration}
The electric connections of the device are illustrated in Fig. \ref{HP_calibration_model}(a). Our device includes four SDMRs. Each SDMR consists of two microrings (marked as MR$_{ij}$, where $i$=1, 2, 3 and 4 indicate the 1st, 2nd, 3rd, and 4th SDMR, respectively; $j$=1 and 2 indicate the 1st and 2nd MRs of SDMR). Here, we define $\text{U}_{i1}$ and $\text{U}_{i2}$ as the voltages applied to 1st and 2nd MRs of the ith SDMR, respectively. PD$_i$ means the photodtector of the ith SDMR. I$_{\text{PD}i}$ is the current read from the PD$_i$. Fig. \ref{HP_calibration_model}(b) presents the experimental flowchart that finding the corresponding ($\text{U}_{i1}$, $\text{U}_{i2}$) for each spectral channel: 

(1) Set the input wavelength as 1530 nm by a tunable laser;

(2) Sweep the input power of $\text{U}_{i1}$ and $\text{U}_{i2}$ from 0 to 70 mW and from 0 to 100 mW, respectively. The size of the total points ($\text{U}_{i1}$, $\text{U}_{i2}$) is 71 $\times$ 101;

(3) For each ($\text{U}_{i1}$, $\text{U}_{i2}$), the photocurrent I$_{\text{PD}i}$ is read;

(4) After all the permutations of $\text{U}_{i1}$ and $\text{U}_{i2}$ is swept, we find the maximum I$_{\text{PD}i}$ and record the corresponding ($\text{U}_{i1}$, $\text{U}_{i2}$);

(5) Increase the wavelength and sweeping the input power of $\text{U}_{i1}$ and $\text{U}_{i2}$ again until 1579 nm.

\begin{figure} [ht]
\centering\includegraphics[width=130mm]{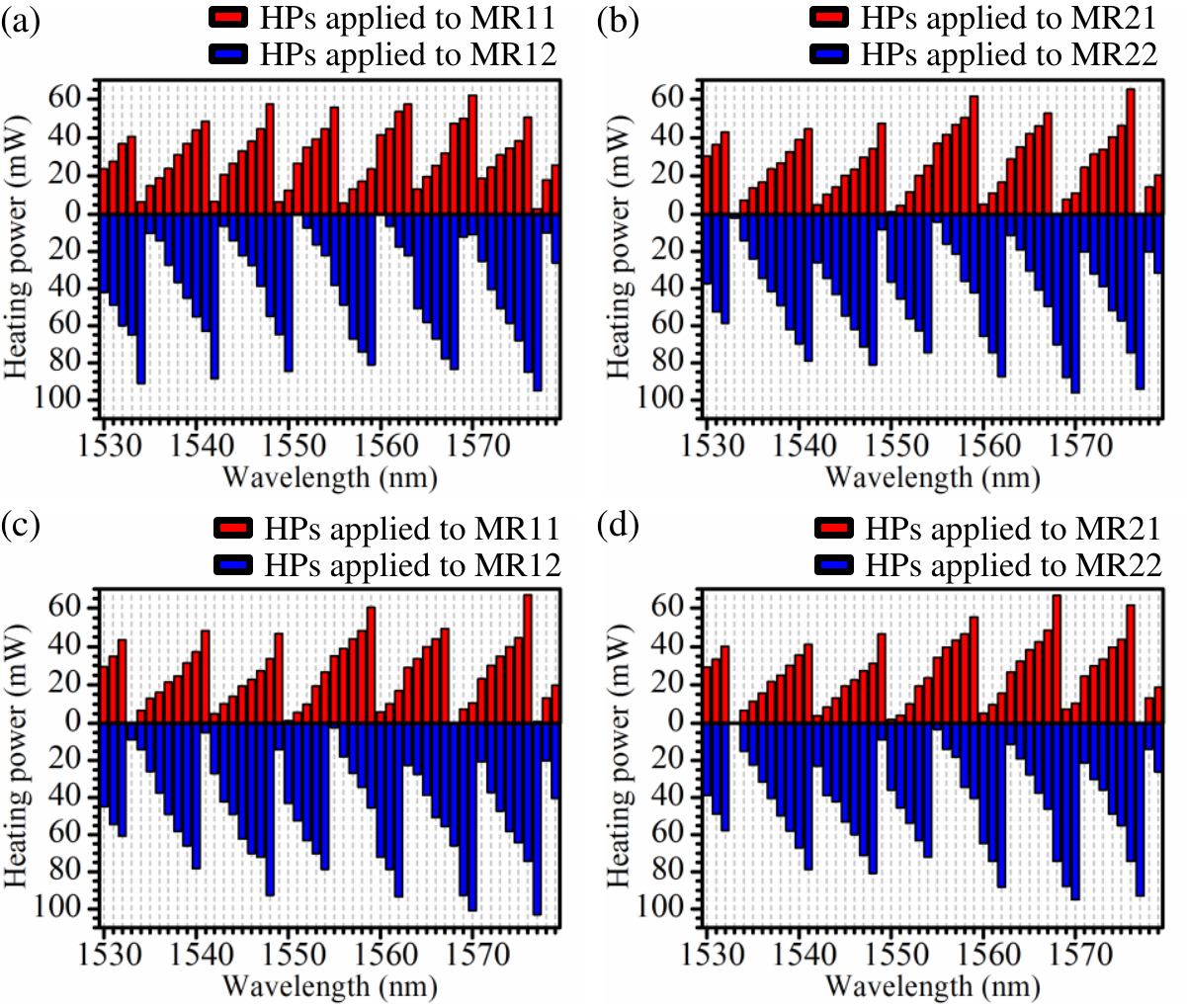}
\caption{ (a)-(d) are the calibrated heating power of MRi1 (red) and MRi2 (blue) for each spectral channel.}
\label{HP_calibration_result}
\end{figure}

The calibration results of the four SDMRs are shown in Fig. \ref{HP_calibration_result}.

\section{Appendix E: Experiment setup}
\begin{figure} [ht]
\centering\includegraphics[width=100mm]{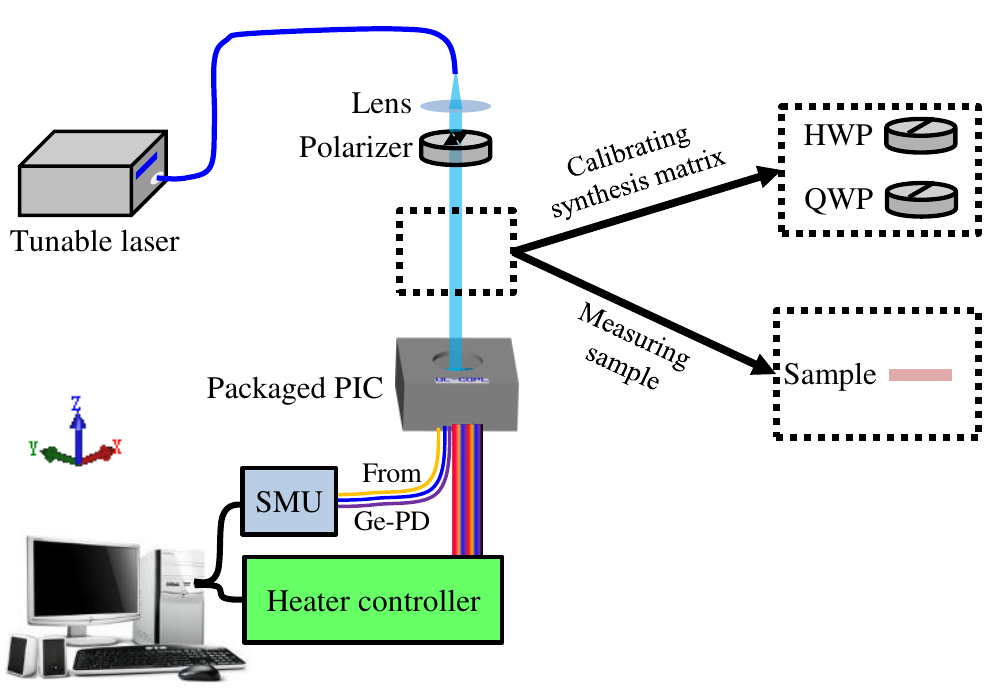}
\caption{ Experiment setup for calibrating the synthesis matrix or characterizing a chiral material. HWP: half-wave plate; QWP: quarter-wave plate. SMU: source measure unit used for reading the current from photodetector.}
\label{Experiment_setup}
\end{figure}

The experiment setup for calibrating the synthesis matrix or characterizing a chiral material is shown in Fig. \ref{Experiment_setup}. The input polarization is controlled by rotating the angles of the half-wave plate (HWP) and quarter-wave plate (QWP) and can be calculated by the angles of HWP and QWP. Therefore, we can calibrate the synthesis matrix of our device using HWP and QWP. If we want to measure the sample, we can replace the HWP and QWP by the sample.   

\section{Appendix F: Materials and Methods}
\subsection{Device Fabrication.} The device was fabricated using a commercial CMOS-compatible SOI process with 193 nm deep-UV lithography at IME, Singapore. The devices were subsequently packaged at our lab. The electrical connections were realized using Westbond’s 7400A Wire Bonder. The plastic cover shell was fabricated using a 3D printer (Ultimaker S5).

\subsection{Chiral Molecule Sample Fabrication.} The CLC material used was a mixture of commercially available Nematic Liquid Crystal (NLC) 20608 (Qingdao Chemicals) and the chiral molecule CB15 (Merck). We have adjusted their ratio (75:25 wt $\%$ ratio), so that we can obtained a CLC with selective reflection band in the near IR region. The CLC mixture was heated above the clearing point (isotropic phase transition) and filled into the LC cell of 9.6 $\mu$m thickness by capillary method and then was slowly cooled down to the room temperature. The cell consists of two indium tin oxide /ITO/ coated transparent glass substrates, which are coated with alignment layers that align CLC molecules parallel to the surface of the substrates.

\subsection{Device characterization.} The calibration of the HP described in the main text was performed using a tunable laser source (Agilent 81600B) with optical power around 3 dBm. The photocurrents from Ge-PD were read by a Keithley 2612B souremeter. The HPs of the heaters were driven using a Keysight E3631A power supply. The light from a high-power wide-band Erbium ASE source (INO) was used to characterize the Si-DMRS. A commercial optical spectrum analyzer (OSA, Yokogawa AQ6370D) was used to measure its spectrum. The synthesis matrix of the proposed spectropolarimeter was calibrated by a polarizer (650-2000 nm, Thorlabs), an HWP (1550 nm, Thorlabs), and a quarter-wave plate (QWP, 1550 nm, Thorlabs). Two stepper motor rotators (K10CR1/M, Thorlabs) were used to control separately the angles of the HWP and QWP.

\section*{Disclosures}

The authors declare no conflicts of interest


\bibliography{OSA-template.bib}

\end{document}